\documentclass[prd,aps,floats,showpacs,twocolumn,twoside,preprintnumbers,superscriptaddress,floatfix]{revtex4}
\usepackage{graphicx}
\usepackage{amsmath}
\usepackage{slashed}
\usepackage{color}
\usepackage{balance}

\begin{document}

%\title{Phase diagram of a three-flavor color superconducting PNJL model}
\title{Universal limiting  pressure for a three-flavor color superconducting PNJL model phase diagram}

\author{A.~Ayriyan}
\email{ayriyan@jinr.ru}
\affiliation{Laboratory of Information Technologies, JINR Dubna, RU-141980 Dubna, 
Russia}

\author{J.~Berdermann}
\email{jens.berdermann@gmail.com}
%\affiliation{German Aerospace, Neustrelitz, Germany}
\affiliation{Deutsches Zentrum f\"ur Luft- und Raumfahrt (DLR), DE-17235 Neustrelitz, Germany}

\author{D.~Blaschke}
\email{blaschke@ift.uni.wroc.pl}
\affiliation{Instytut Fizyki Teoretycznej, Uniwersytet Wroc{\l}awski, 
PL-50-204 Wroc{\l}aw, Poland}
\affiliation{Bogoliubov Laboratory of Theoretical Physics, JINR Dubna, RU-141980 Dubna, Russia}
\affiliation{National Research Nuclear University (MEPhI), RU-115409 Moscow, Russia}

\author{R.~{\L}astowiecki}
\email{lastowiecki@ift.uni.wroc.pl}
\affiliation{Instytut Fizyki Teoretycznej, Uniwersytet Wroc{\l}awski, 
PL-50-204 Wroc{\l}aw, Poland}

\date{\today}

\begin{abstract}
The phase diagram of a three-flavor Polyakov-loop Nambu-Jona-Lasinio model 
is analyzed for the case of isospin symmetric matter with color superconducting phases.
The coexistence of chiral symmetry breaking and two-flavor color superconductivity (2SC phase) 
and a thermodynamic instability due to the implementation of a color neutrality constraint is observed.
It is suggested to use a universal hadronization pressure to estimate the phase border between 
hadronic and quark-gluon plasma phases.
Trajectories of constant entropy per baryon are analyzed for conditions appropriate 
for heavy-ion collisions in the NICA-FAIR energy range.  
\end{abstract}
\maketitle

\section{Introduction}
The phase diagram of quantum chromodynamics (QCD) is in the center of 
attention of modern physics~\cite{BraunMunzinger:2008tz}. 
It is an object of intensive studies at already existing (LHC, RHIC, SPS, SIS) and 
planned (FAIR-CBM, NICA-MPD) heavy-ion collision experiments \cite{Braun-Munzinger:2014pya}.
In addition, the high density part of the QCD phase 
diagram is of interest for astrophysics, in particular for scenarios of 
supernova explosions~\cite{Sagert:2008ka,Fischer:2011zj}, 
neutron star (NS) mergers~\cite{Rezzolla:2010,Bauswein:2012ya,Takami:2014zpa}, 
and NS structure~\cite{Glendenning:1997wn,Blaschke:2001uj,Haensel:2007yy}.
For QCD motivated hybrid star models that fulfil the $2M_\odot$ constraint 
\cite{Demorest:2010bx,Fonseca:2016tux,Antoniadis:2013pzd}
on the maximum mass, see for instance 
\cite{Alford:2006vz,Klahn:2006iw,Klahn:2013kga,Kojo:2014rca}
and references therein.
A set of constraints for the high-density equation of state (EoS) from the phenomenology of 
compact stars and heavy-ion collisions has been compiled in \cite{Klahn:2006ir}
and applied also to the case of hybrid EoS in \cite{Klahn:2011au}.
Of special interest are statements about the possible existence of a critical endpoint 
in the QCD phase diagram \cite{Hatsuda:2006ps,Powell:2011ig}
that could eventually be supported by NS phenomenology
\cite{Blaschke:2013ana,Benic:2014jia,Ayriyan:2014nua,Alvarez-Castillo:2016oln}
or in black-hole formation \cite{Ohnishi:2011jv}.

From the theoretical standpoint the high temperature region of phase diagram 
can be accessed by lattice simulations \cite{Borsanyi:2013bia,Bazavov:2014pvz}.
Unfortunately, at finite chemical potential direct Monte-Carlo calculations of QCD on the
lattice run into the, yet to be solved, sign problem (see, e.g., Ref.~\cite{Karsch:1998jm} for an overview).
In light of these difficulties effective models have been developed and applied 
in order to extract information from laboratory experiments and compact star astrophysics as well as to 
cast predictions on the phase structure of QCD at high densities 
(see, e.g., Ref.~\cite{Friman:2011zz} for an introduction).
Most popular for calculating the quark matter equation of state and for the study of the QCD phase diagram  
became the class of so-called chiral quark models built in analogy to the Nambu--Jona-Lasinio (NJL) 
model~\cite{Nambu:1961tp}; for reviews see, e.g., Refs.~\cite{Klevansky:1992qe,Hatsuda:1994,Buballa:2003qv,Fukushima:2010bq,Fukushima:2013rx}.

While being successful in capturing aspects of chiral symmetry breaking and restoration as well as pion 
properties in hot, dense quark matter, the main shortcoming of these models is the lack of a confinement mechanism.  
A partial fix to this problem is provided by the coupling of the chiral quark sector to the Polyakov-loop with an appropriate temperature dependent effective potential fitted to pure gauge lattice QCD thermodynamics results at~$\mu=0$~\cite{Ratti:2005jh}, which is then extended also to finite chemical potentials, e.g., in Refs.~\cite{Roessner:2006xn,Blaschke:2010ka}.

It has been shown \cite{Schafer:1999jg} that the QCD Lagrangian at asymptotically high chemical 
potentials exhibits the Cooper instability \cite{Cooper:1956zz}, due to attractive 
quark-quark interaction in the color anti-triplet channel.
In the framework of NJL-like models the question of the color superconductivity
(CS) resulting from the BCS mechanism has been studied and
a large variety of phases has been proposed \cite{Alford:2001}, 
including anisotropic phases leading to a color superconducting crystalline 
phases, see \cite{Alford:2007xm,Anglani:2013gfu} for recent reviews.

The analysis of the PNJL model phase diagram with color superconductivity has been 
performed for the two-flavor system in \cite{GomezDumm:2008sk}.
The aim of this paper is to extend this study to three-flavor quark matter.
The phase diagram of the NJL model, without the coupling to the Polyakov loop variable, has been 
studied for a certain parameter choice \cite{Blaschke:2005uj,Ruester:2005jc}
and the possibility of neutrino trapping with its effect on the phase diagram has been considered in 
\cite{Ruester:2005ib,Sandin:2007zr}.
In any model for high-density matter, the $2M_\odot$ mass constraints for neutron stars 
%on the model parameters following from the mass measurements of PSR~J1614-2230~\cite{Demorest:2010bx,Fonseca:2016tux} and PSR~J0348-0432~\cite{Antoniadis:2013pzd} 
has to be taken into account (for details see \cite{Klahn:2013kga}).

\section{Color-superconducting Polyakov-Nambu-Jona-Lasinio model}
In the present study we use a state-of-the-art PNJL model including 
scalar, diquark and vector interaction channels as well as the Polyakov-loop variable.
The~Kobayashi-Maskawa-'t Hooft (KMT) determinant interaction term has been 
considered in \cite{Blaschke:2010vj} in the context of this model and the $2M_\odot$ constraint
from PSR~J1614-2230 but will be suppressed for clarity of the present study.

The Lagrangian of the model can be decomposed into three parts
\begin{equation}
 {\cal L} = {\cal L}_{0} + {\cal L}_{\rm int} + {\cal U}(T, \Phi)~,
\end{equation}
with the free Dirac part 
\begin{equation}
 {\cal L}_{0} = \bar{q}(-i\gamma_\mu D_\mu + \hat{m} + \hat{\mu})q~,
\end{equation}
where $\hat{m}={\rm diag}(m_u,m_d,m_s)$ and $\hat{\mu}={\rm diag}(\mu_u,\mu_d,\mu_s)$. 
The interaction part given as
\begin{eqnarray}
{\cal{L}}_{\rm int} &=& 
G_S\sum_{a=0}^{8}\left[(\bar{q}\tau_a q)^2 
+ \eta_V(\bar{q}i\gamma_0q)^2\right] 
\nonumber \\ 
&+& 
G_S\eta_D\hspace{-4mm}\sum_{a,b=2,5,7}
(\bar{q}i\gamma_5 \tau_a \lambda_b C \bar{q}^T)(q^TCi\gamma_5\tau_a\lambda_a q).
\end{eqnarray}
The covariant derivative $D_\mu= \partial_\mu - i\bar{A}_\mu$ 
includes minimal coupling of quark fields to 
the background gluon field $G^\mu_a$
%\begin{equation}
% D_\mu = \partial_\mu - iA_\mu.
%\end{equation}
which is assumed in the form
\begin{equation}
 \bar{A}_\mu = g\delta_{\mu 0}G^\mu_a\lambda^a/2.
\end{equation}
The traced Polyakov loop can be expressed as
\begin{equation}
 \Phi = \frac{1}{3}Tr\exp(i\beta\phi),
\end{equation}
where $\phi = i\bar{A}_0$.
In Polyakov gauge the matrix $\phi$ is
\begin{equation}
 \phi = \phi_3\lambda_3 + \phi_8\lambda_8,
\end{equation}
with two independent variables $\phi_3$ and $\phi_8$.
The Polyakov-loop potential is chosen in the logarithmic form
\begin{eqnarray}
\frac{{\cal U}(\Phi, T)}{T^4} &=& -\frac{1}{2}a(T)\Phi^\star\Phi\quad\qquad\qquad\qquad\qquad\qquad\qquad\nonumber\\
&&+ b(T)\ln\left[1 - 6\Phi^\star\Phi + 4(\Phi^3 + {\Phi^\star}^3)\right.\nonumber \\
&&\left.- 3(\Phi^\star\Phi)^2\right]~,
\end{eqnarray}
with coefficients $a(T)$ and $b(T)$ taken from \cite{Roessner:2006xn}
$$
a(T)=a_0+a_1\left(\frac{T_0}{T}\right)+a_2\left(\frac{T_0}{T}\right)^2, \quad
b(T)=b_3\left(\frac{T_0}{T}\right)^3~,
$$
where $a_0 = 3.51$, $a_1 = -2.47$, $a_2 = 15.2$ and $b_3 = -1.75$.

The parameters of the model are set in vacuum, by fitting meson properties, 
in particular the pion mass, pion decay constant and kaon mass.
The NJL model contains four parameters: the cutoff for the divergent three-momentum integrals,
the scalar coupling constant, the current mass of light quarks (in principle those are two parameters,
but for simplicity we can choose to have $m_u = m_d$) and the current mass of strange quarks $m_s$.
This leaves one parameter which is fitted with either quark condensate in vacuum or to the 
constituent quark mass.
This leaves some freedom in setting the value of the scalar coupling strength.
In this work we will choose one of parametrisations compatible with observables,
to be exact we chose parametrization with constituent quark mass $M(p=0) = 367.5$~MeV
from Ref.~\cite{Grigorian:2006qe}, where the value of the strange quark mass has been corrected \cite{Sandin}.
The resulting parameter values are
\begin{eqnarray}
\Lambda &=& 602.3\nonumber\\
m_{u,d} &=& 5.5\nonumber\\
m_s &=& 138.757\nonumber\\
G_S &=& 6.38427\nonumber.
\end{eqnarray}

The Lagrangian can be used to obtain the mean-field approximation to the grand canonical thermodynamical potential
\begin{eqnarray}
\Omega\left(T,\{\mu\}\right) &=& \frac{\phi_u^2 + \phi_d^2 + \phi^2_s}{8 G_S} 
- \frac{\omega^2_u + \omega^2_d + \omega^2_s}{8 G_V} \qquad\qquad\qquad\nonumber \\
&+& \frac{\Delta^2_{ud} + \Delta^2_{us} + \Delta^2_{ds}}{4G_D} \nonumber\\
&-& \int{\frac{d^3p}{(2\pi)^3}}\sum_{n=1}^{18}\left[E_n + 
2T \ln\left(1 + e^{-E_n/T}\right)\right] \nonumber\\
&+& \Omega_{lep} - \Omega_0 + {\cal U}(\Phi,T)~.
\end{eqnarray}

We are interested in the mean-field equilibrium value of $\Omega$.
In order to reach that goal it is necessary to find the extremum of $\Omega$ with 
respect to all order parameters.
This procedure results in a set of coupled gap equations
%I think, this record is not correct. Sasha
%\begin{equation}
%\frac{\partial\Omega(T,\mu)}{(\partial\phi_f,\partial\Delta_{ik},\partial\phi_3,\partial\phi_8)} = 0
%\end{equation}
\begin{equation}
\frac{\partial\Omega}{\partial\phi_f}=
\frac{\partial\Omega}{\partial\Delta_{ik}}=
\frac{\partial\Omega}{\partial\phi_3}=
\frac{\partial\Omega}{\partial\phi_8}=0~,
\end{equation}
that need to be solved self-consistently.
Additionally color-neutrality should be observed, which leads to additional 
equations for color chemical potentials $\mu_3$ and~$\mu_8$, see 
\cite{Buballa:2003qv,GomezDumm:2008sk}

\begin{equation}
\frac{\partial\Omega}{\partial\mu_3}=
\frac{\partial\Omega}{\partial\mu_8}=0~.
\end{equation}

The set of values of the free parameters $\eta_D$ and $\eta_V$ has been chosen to fulfil the PSR J1614-2230 and PSR~J0348-0432 mass constraint as 
$\eta_D = 1.0$ and $\eta_V = 0.30$ \cite{Klahn:2013kga}.
%\begin{table}[htb]
%\begin{tabular}{cc}
%\qquad $\eta_D$ \qquad & \qquad $\eta_V$ \qquad\\
%\qquad 1.0 \qquad & \qquad 0.30 \qquad\\%\hline
%\end{tabular}
%\end{table}

\begin{figure*}[!th]
	\begin{tabular}{cc}
		\includegraphics[width=7.5cm]{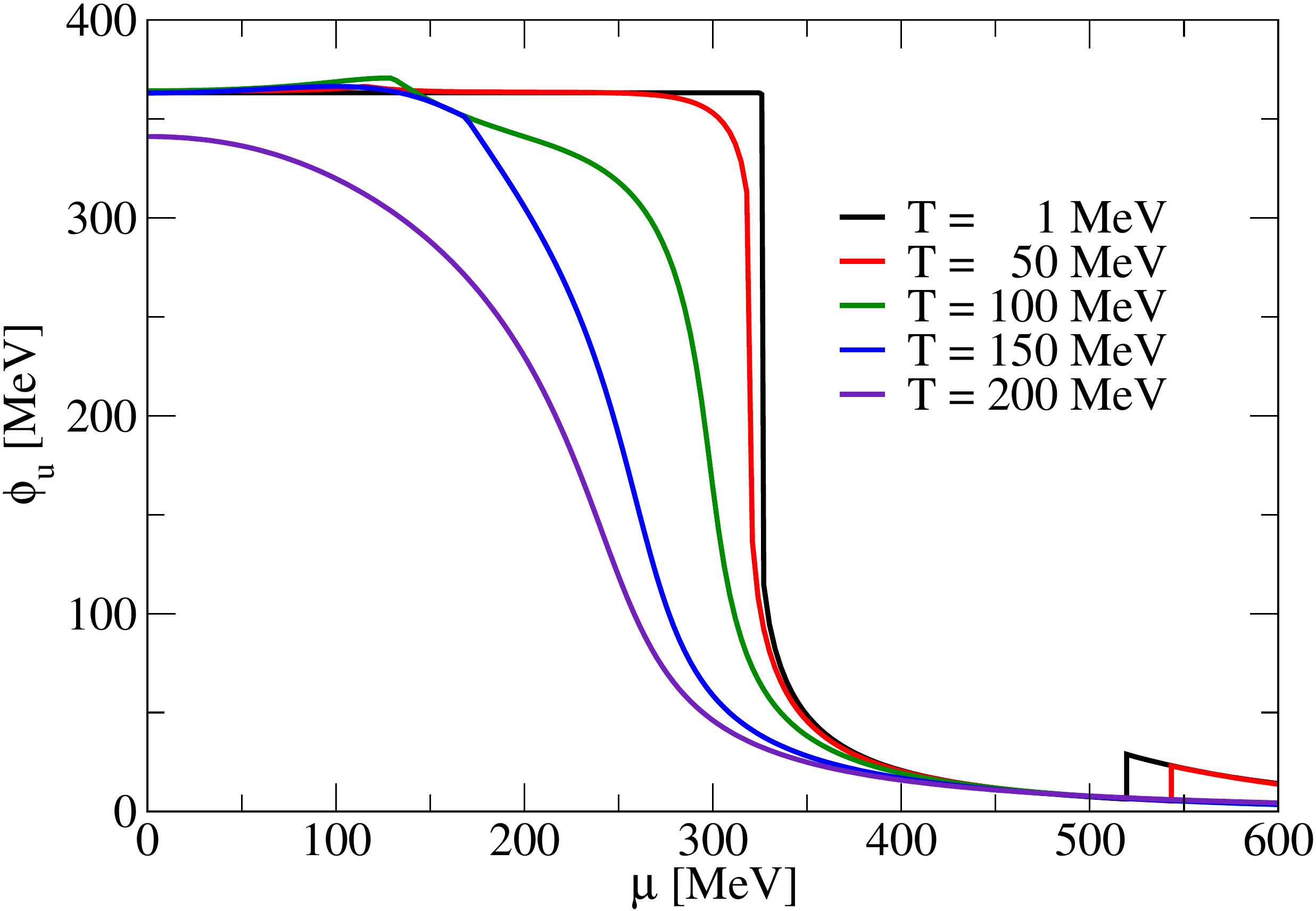}
		&
		\includegraphics[width=7.5cm]{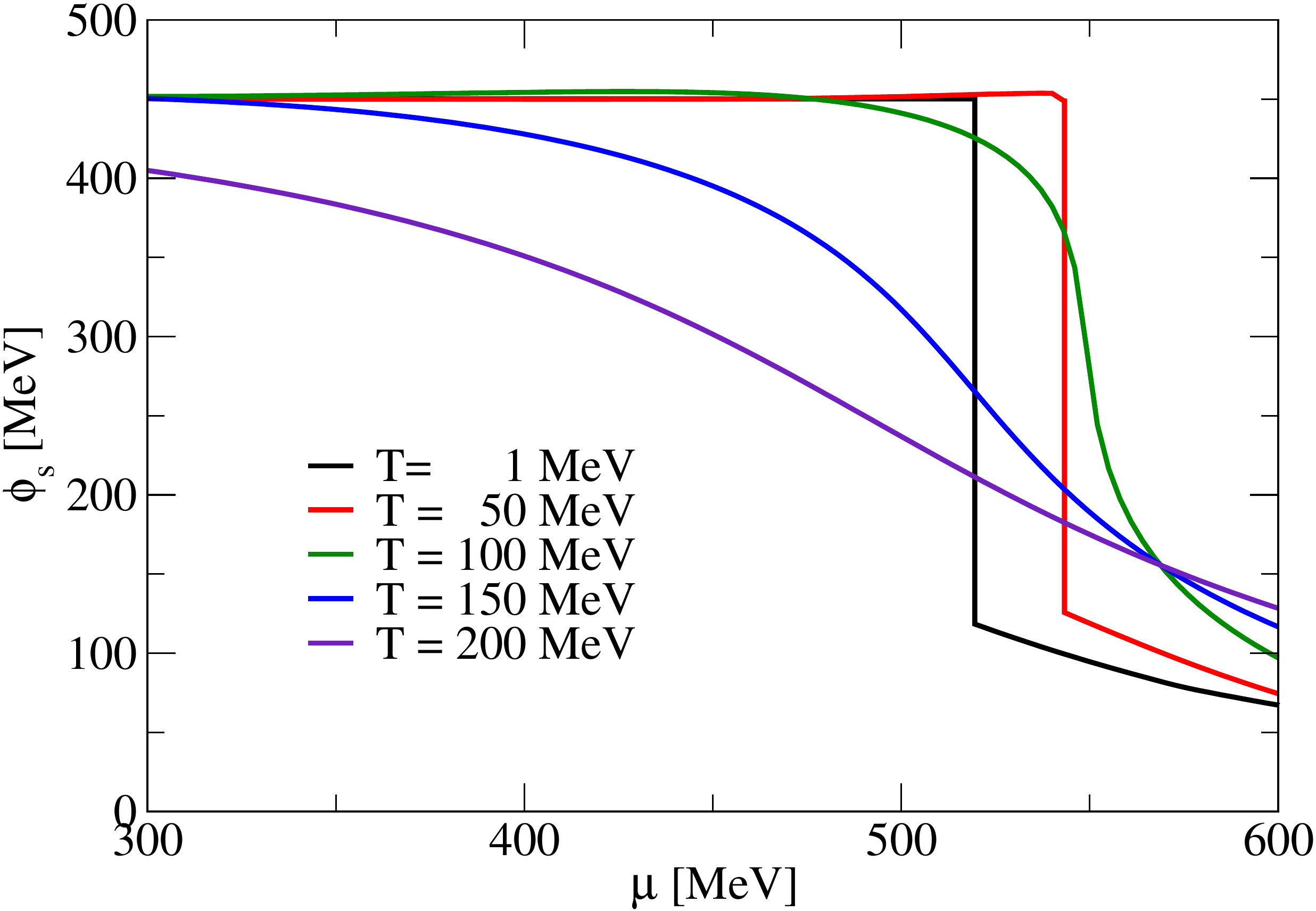}	
		\\
		\includegraphics[width=7.5cm]{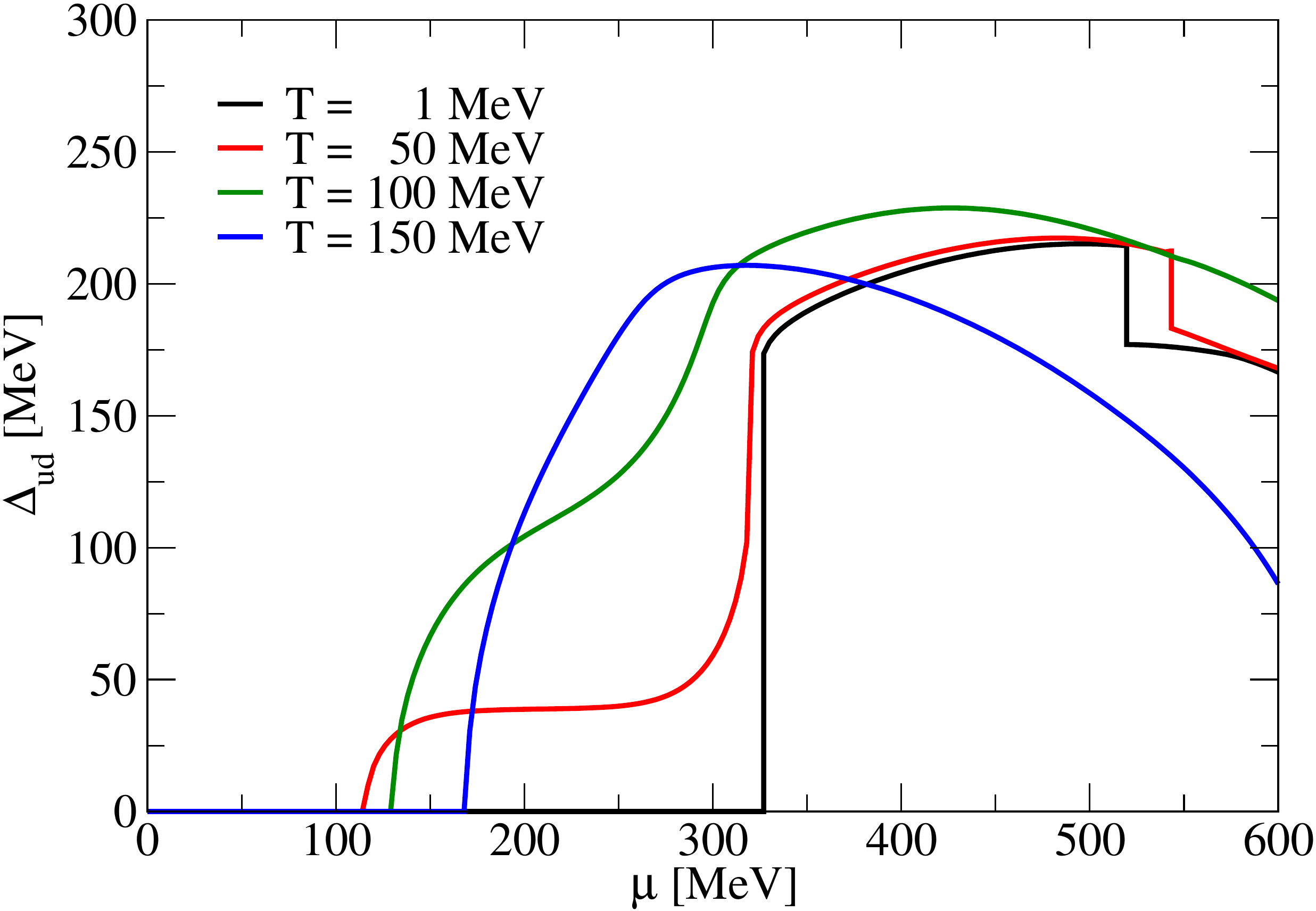}
		&		
		\includegraphics[width=7.5cm]{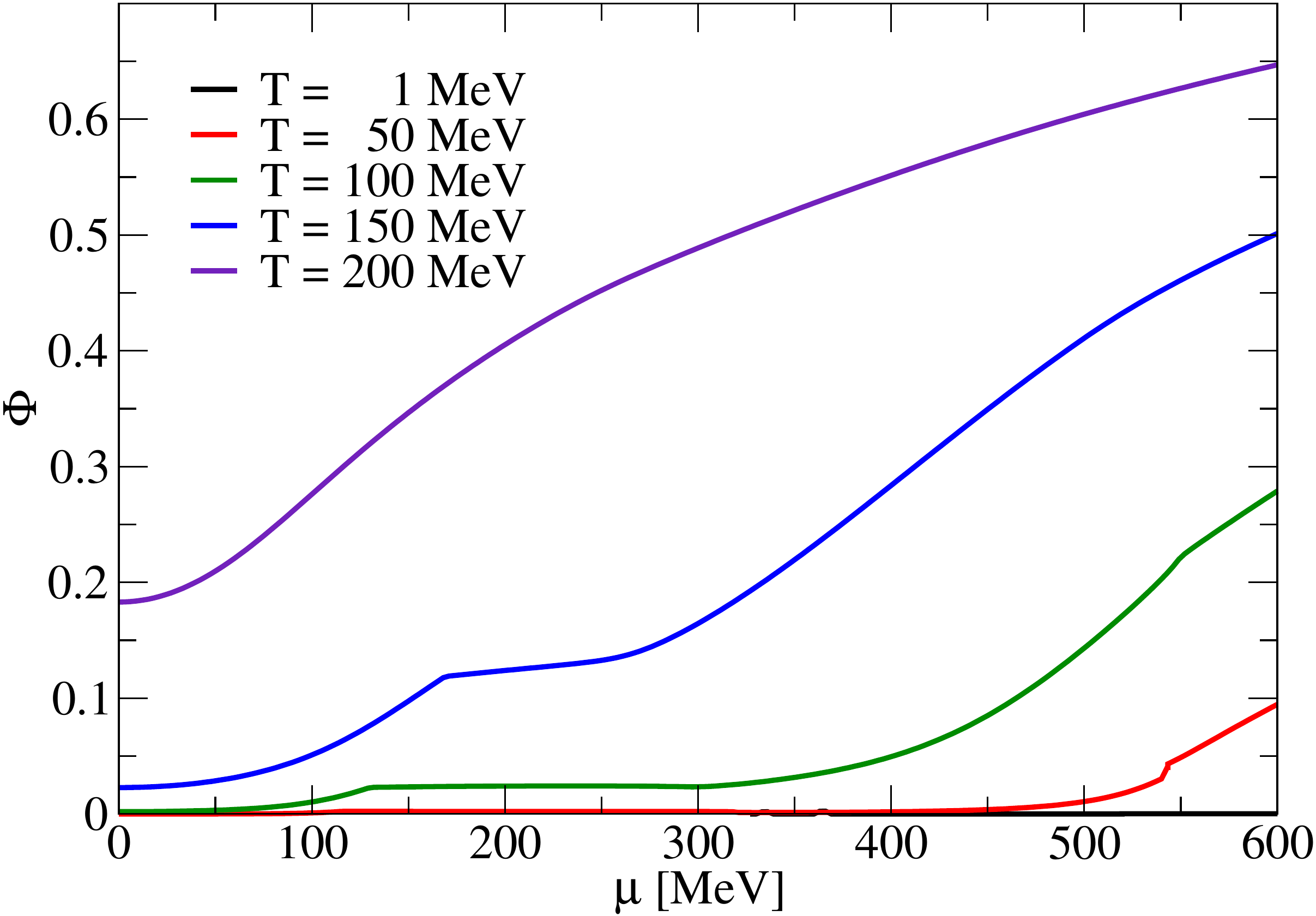}
	\end{tabular}
	\caption{(Color online) Solutions of the gap equations for isospin symmetric quark matter as functions of the chemical potenial $\mu$ for selected temperatures $T=1$, 50, 100, 150 and 200 MeV.  
	The behavior of the light quark mass gap $\phi_u=\phi_d$ (upper left panel) signals a first-order chiral transition at low and a crossover higher temperatures with a critical endpoint between $T=50$ and 100 MeV. The mass gap for strange quarks $\phi_s$ is shown in the upper left panel and indicates a sequential chiral restoration at higher chemical potentials than for light quarks. 
	The color antitriplet pairing gap between up and down quarks ($\Delta_{ud}$) is shown in the lower left panel and the traced Polyakov loop $\Phi$ in the lower right panel.
	 \label{Fig:Condensates}
}
\end{figure*}

\section{Results for isospin symmetric matter}
In heavy-ion collisions matter is to a very good approximation in an isospin symmetric state.
This implies the following relation for the chemical potentials of quark flavors
\begin{equation}
 \mu_u = \mu_d, \qquad \mu_s \simeq 0.
\end{equation}
No electrons are present during the collision and the system is not 
electrically neutral.
For the purpose of this paper we will restrict our discussion to this case.
We should note, however, that for astrophysical applications more relevant 
is the state of matter in beta equilibrium.

We begin the discussion of the results with a close look at the behavior of the mean field condensates
and the Polyakov loop variable.
In Fig.~\ref{Fig:Condensates} we present the mass gaps, the $ud$-pairing gap and the Polyakov loop as a function of the chemical potential for selected temperatures.

For the mass gap of light quarks $\phi_u=\phi_d$ (shown in the upper left panel) we observe a standard behavior: at small temperatures the chiral symmetry restoration is a first order phase transition with an apparent jump in the condensate.
When the temperature is increasing the transition tends to be more and more of a crossover type,
with gradual decrease in the condensate.
The chemical potential dependence of the strange quark mass gap $\phi_s$ (shown in the upper right panel) is similar:  with increasing temperature, the first order transition changes to a crossover. 
Only the magnitude of the gap and the chemical potential where the transition takes place is larger than in the light quark sector. The transitions occur sequentially with increasing baryon number density.  

The pairing gap of up and down quarks $\Delta_{ud}$ is shown in the lower left panel of 
Fig.~\ref{Fig:Condensates}. 
For the temperature $T=1$ MeV (which is equivalent to the zero temperature case)
the behavior is typical for the NJL model: the $ud$ pairing gap gets switched on with a jump at the same critical chemical potential where the light quark mass gap jumps down by one order of magnitude. 
Chiral symmetry breaking and color superconductivity are anticorrelated.
Because of the still too big mass difference between light and strange quarks, no pair condensation in the light-strange mixed pairing channels occurs,   $\Delta_{us}=\Delta_{ds}=0$, which characterizes the 
two-flavor color superconducting (2SC) phase.
At chemical potentials around $520$ MeV a second, smaller jump occurs which is connected to the partial chiral restoration in the strange quark sector (see the upper right panel) which entails the appearance of the color-flavor locking (CFL) phase characterized by three nonvanishing pairing gaps instead of one, 
$\Delta_{ud}\neq 0$, $\Delta_{us}\neq 0$, $\Delta_{ds}\neq 0$.

In the present model the Polyakov loop has been included as a way to account to some
degree for the confinement of quarks and gluons.
We remind that the Polyakov loop equal to zero signals confinement,
while its approach to one indicates deconfinement.
In the lower right panel its chemical potential dependence is shown. 
It is close to zero for small temperatures and chemical potentials and rises as a function of both.

An interesting phenomenon is observed when the temperature is increased. 
The pairing gap develops non-zero values even at chemical potentials well below the (pseudo-)critical one for chiral restoration so that a phase coexistence of chiral symmetry breaking and color superconductivity appears. 
It is in particular in this region of the $T-\mu$ plane that negative pressure and negative baryon density occur, signalling a thermodynamic instability.
We suspect that this is a consequence of the procedure for establishing color neutrality of the system, as defined by a vanishing of the color densities $n_3=n_r-n_g$ and $n_8=(n_r+n_g-2n_8)/\sqrt{3}$ which are conjugate to the color chemical potentials $\mu_3=\mu_r - \mu_g$ and $\mu_8=(\mu_r+\mu_g-2\mu_b)/\sqrt{3}$, respectively (see chapter 6 of \cite{Buballa:2003qv} for details).
The shortcomings of this definition of color neutrality in the presence of a 2SC diquark condensate have been discussed in \cite{Buballa:2005bv}, see also \cite{He:2005jq,Blaschke:2005km} for the NJL model.
In the presence of a Polyakov-loop field the situation with respect to color neutrality is further obscured and requires particular attention, see  \cite{Abuki:2009dt} for a special model realization.

It is clear that in a quark matter model which would realize quark and gluon confinement in color singlet states in the hadronic phase no problems with color neutrality and thermodynamic stability shall occur in that region since it would be covered by the dominance of hadronic degrees of freedom.

\begin{figure}[!th]
		\includegraphics[width=0.49\textwidth]{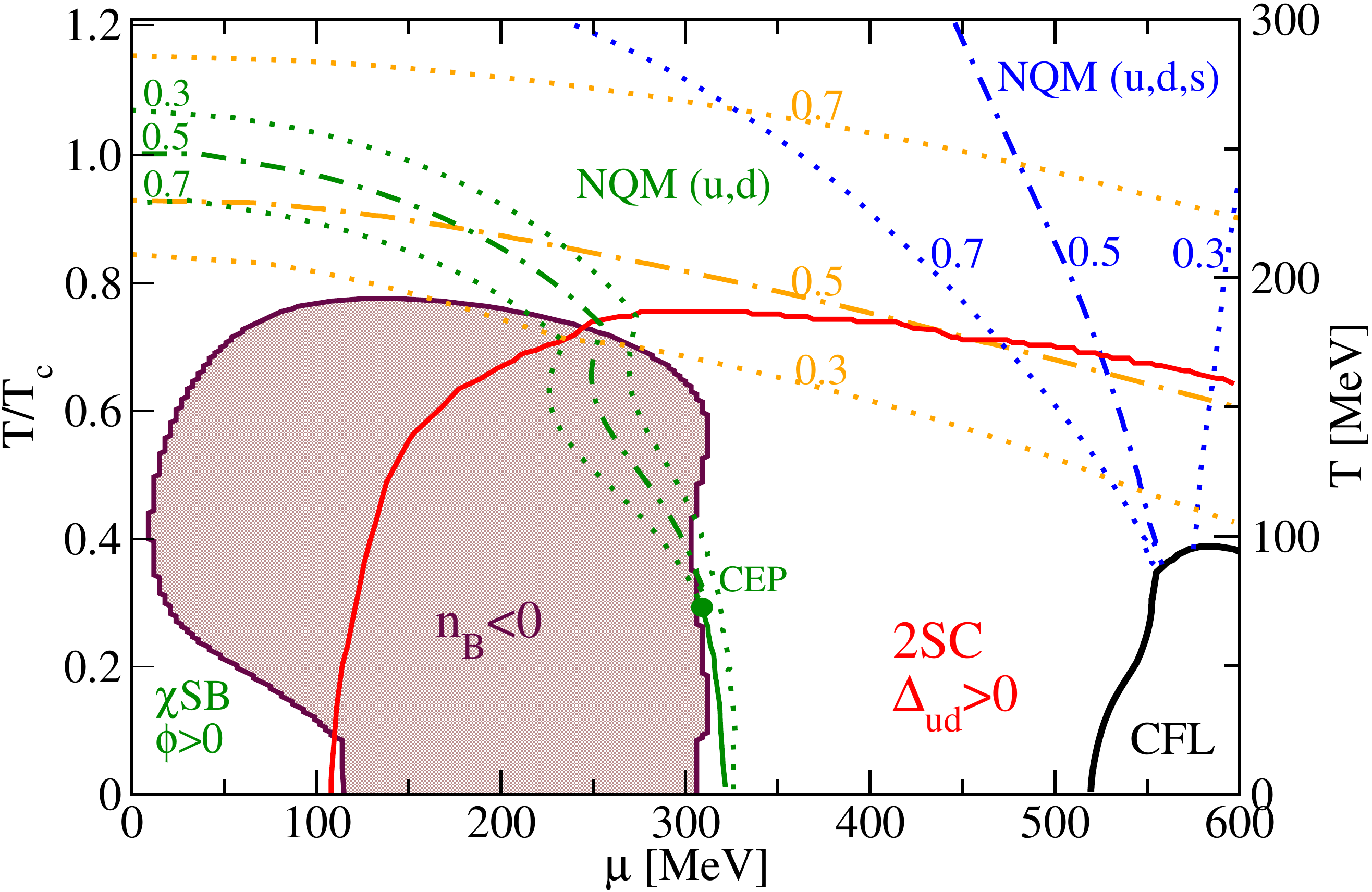}
	\caption{(Color online) Phase diagram of symmetric matter in the 3FCS PNJL model with lines of constant values for the order parameters: vanishing 2SC (red) and CFL (black) pairing gaps, Polyakov loop (orange), light quark mass (green) and strange quark mass (blue) relative to their vacuum values. 
	The instability region ($n_B<0$) is shown as filled with brown color.
	 \label{Fig:MainPlot}
}
	\end{figure}
	
The results of this work are summed up in Fig.~\ref{Fig:MainPlot}.
The phase structure of the model (note that no phase transition to quark matter has been constructed here)
consists of a chirally broken phase, the previously mentioned region of instability with negative density and pressure, chirally restored normal quark matter for high temperatures, the 2SC phase at moderate chemical potentials and not too high temperatures (which in our case penetrates also to the region where chiral symmetry is not yet restored) and finally the CFL phase superconductor at low temperatures and high chemical potentials.

\begin{figure}[!th]
		\includegraphics[width=0.48\textwidth]{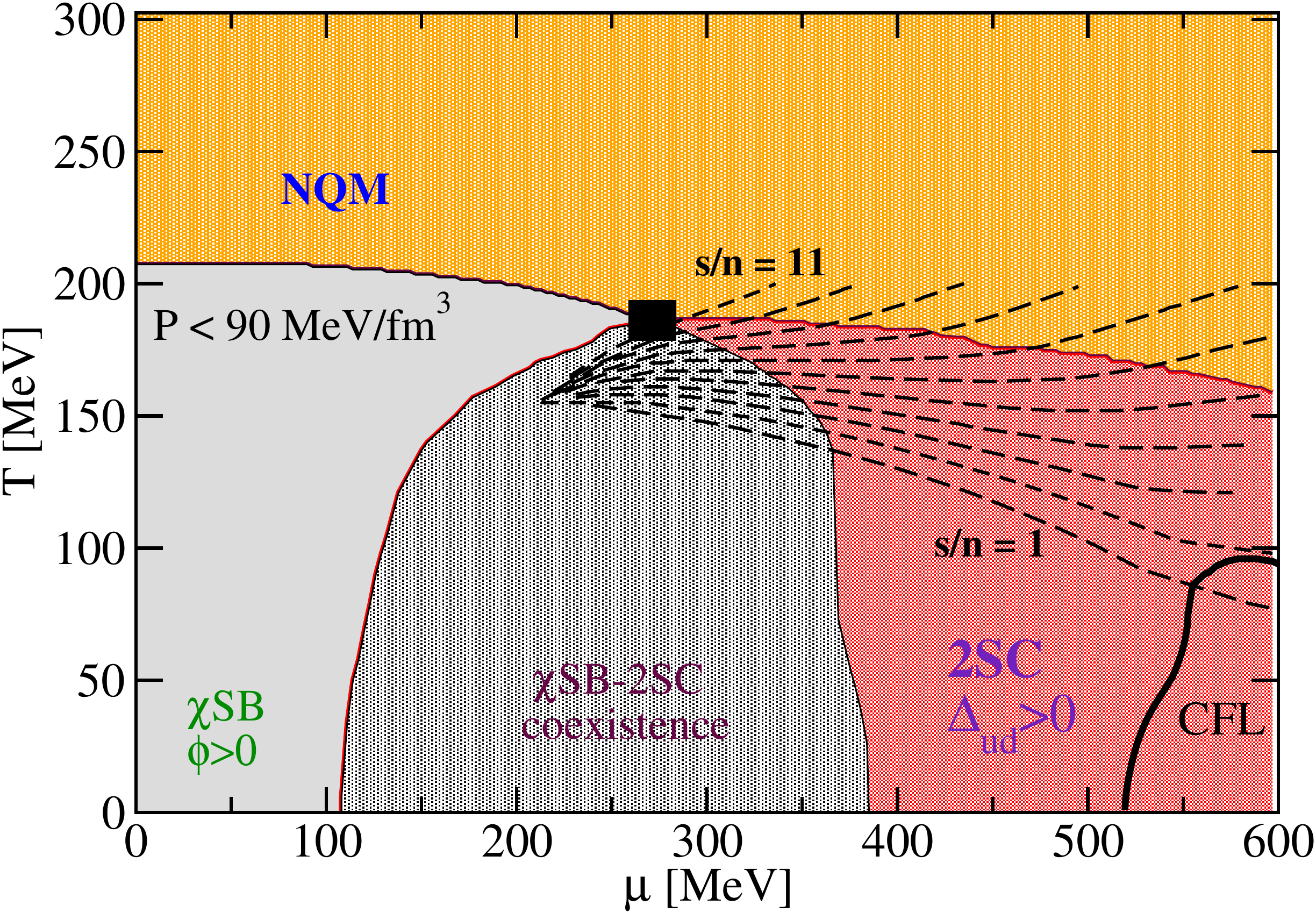}
	\caption{(Color online) Phase diagram of the color superconducting three-flavor PNJL model with the hadronic phase region defined by a universal hadronization pressure~\cite{Petran:2013qla} of $90~$MeV/fm$^3$.
	 \label{Fig:PhaseDiagram}
}
\end{figure}

In Fig.~\ref{Fig:PhaseDiagram} we extract a schematic phase diagram for low-energy QCD from the results of the color superconducting three-flavour PNJL model and its order parameters where we identify the region of the hadronic phase following the reasoning of Ref.~\cite{Petran:2013qla} where the concept of a universal hadronisation pressure has been introduced on the basis of a statistical model parametrization of particle abundances at chemical freeze-out.
 
For the discussion of potential applications of our results  to heavy ion collisions we plotted lines of constant entropy per baryon ranging from ${s}/{n} = 1$ up to ${s}/{n} = 11$.
This could be approximate trajectories of the evolution of a fireball in HIC for the lower range of NICA and FAIR energies, see Fig.~2 of Ref.~\cite{Ivanov:2016hes}.
In this figure we include also the line of constant pressure equal to $82 \pm 8~$MeV/fm$^3$ 
(we take the upper limit of $90~$MeV/fm$^3$) to estimate the phase border between the quark-gluon 
plasma phases and the hadronic phases of strongly interacting matter.

We note here that the previously mentioned instability region is covered in the phase diagram
completely by the hadronic phase, or at least is located where we would expect the hadronic degrees of freedom to be the relevant ones.
This ``coincidence'' allows to use the model in the thermodynamical parameters range where it is relevant,
despite the inherent instabilities in low energy region.
The three-flavor color superconducting PNJL model with a phase transition to hadronic matter described by the DD2 model was obtained in \cite{Blaschke:2010ka} by a Maxwell construction.
Indeed the potential instability region is covered there with the hadronic equation of state.
\newline

\section{Summary and Conclusions}

The phase diagram for the three-flavor PNJL model with color superconductivity obtained here has three  
order parameters that characterize the phase structure
\begin{enumerate}
\item chiral condensates (quark masses) - chiral symmetry breaking in light and strange sectors;
\item diquark gaps (color superconductivity) - 2SC and CFL phases;
\item Polyakov loop (confinement) - Z(3) symmetry breaking.
\end{enumerate}
There is a region of instability characterized by a negative pressure that is obtained in the confining phase
where chiral symmetry is broken. 
We exclude this region by defining a hadronic phase with non-negative pressure below 
$P_{\rm crit} \sim 82 \pm 8~$MeV/fm$^3$, the universal hadronization pressure \cite{Petran:2013qla}.
The light and strange quarks appear sequentially in the system.
Lines of constant entropy per baryon indicate paths for the dynamical evolution of a fireball in a heavy-ion collision.
We emphasize the accessibility at NICA/FAIR energies of a "quarkyonic matter" region with partial chiral symmetry restoration inside the hadronic world, as well as a quadruple point (black square) 
\cite{Ayriyan:2016pat} where the four phases meet.

\section*{Acknowledgements}

We acknowledge collaboration with D. Alvarez-Castillo, H. Grigorian, F. Sandin and S. Typel. 
The visits of A.A. at the University of Wroclaw were supported by the Bogoliubov-Infeld program for collaboration between JINR Dubna and Polish Institutes as well as by the COST Action "NewCompStar"
under grant number COST-STSM-MP1304-33291.  
D.B. received funding from NCN under grant number UMO-2011/02/A/ST2/00306.


\begin{thebibliography}{99}

%\cite{BraunMunzinger:2008tz}
\bibitem{BraunMunzinger:2008tz} 
  P.~Braun-Munzinger and J.~Wambach,
  %``The Phase Diagram of Strongly-Interacting Matter,''
  Rev.\ Mod.\ Phys.\  {\bf 81}, 1031 (2009).
%  [arXiv:0801.4256 [hep-ph]].
  %%CITATION = ARXIV:0801.4256;%%

%\cite{Braun-Munzinger:2014pya}
\bibitem{Braun-Munzinger:2014pya} 
  P.~Braun-Munzinger, B.~Friman and J.~Stachel,
  %``Proceedings, 24th International Conference on Ultra-Relativistic Nucleus-Nucleus Collisions (Quark Matter 2014) : Darmstadt, Germany, May 19-24, 2014,''
  Nucl.\ Phys.\ A {\bf 931}, pp.1 (2014).
  %%CITATION = NUPHA,A931,pp.1;%%
  
%\cite{Sagert:2008ka}
\bibitem{Sagert:2008ka} 
  I.~Sagert, T.~Fischer, M.~Hempel, G.~Pagliara, J.~Schaffner-Bielich, A.~Mezzacappa, F.-K.~Thielemann and M.~Liebend\"orfer,
  %``Signals of the QCD phase transition in core-collapse supernovae,''
  Phys.\ Rev.\ Lett.\  {\bf 102}, 081101 (2009).
%  [arXiv:0809.4225 [astro-ph]].
  %%CITATION = ARXIV:0809.4225;%%

%\cite{Fischer:2011zj}
\bibitem{Fischer:2011zj} 
  T.~Fischer, D.~Blaschke, M.~Hempel, T.~Kl\"ahn, R.~{\L}astowiecki, 
  M.~Liebend\"orfer, G.~Martinez-Pinedo and G.~Pagliara {\it et al.},
  %``Core collapse supernovae in the QCD phase diagram,''
  Phys.\ Atom.\ Nucl.\  {\bf 75}, 613 (2012).
  %[arXiv:1103.3004 [astro-ph.HE]].
  %%CITATION = ARXIV:1103.3004;%%

%\cite{Rezzolla:2010}
\bibitem{Rezzolla:2010}
  Bruno~Giacomazzo, Luciano~Rezzolla, Luca~Baiotti, David~Link, and Jos\'{e}~A.~Font,
  %``General Relativistic Simulations of Binary Neutron Star Mergers'',
  AIP Conf. Proc. 1358, pp. 187-190 (2010).

%\cite{Bauswein:2012ya}
\bibitem{Bauswein:2012ya} 
  A.~Bauswein, H.~T.~Janka, K.~Hebeler and A.~Schwenk,
  %``Equation-of-state dependence of the gravitational-wave signal from the ring-down phase of neutron-star mergers,''
  Phys.\ Rev.\ D {\bf 86}, 063001 (2012).
%  [arXiv:1204.1888 [astro-ph.SR]].
  %%CITATION = ARXIV:1204.1888;%%

%\cite{Takami:2014zpa}
\bibitem{Takami:2014zpa} 
  K.~Takami, L.~Rezzolla and L.~Baiotti,
  %``Constraining the Equation of State of Neutron Stars from Binary Mergers,''
  Phys.\ Rev.\ Lett.\  {\bf 113}, no. 9, 091104 (2014).
%  [arXiv:1403.5672 [gr-qc]].
  %%CITATION = ARXIV:1403.5672;%%

%\cite{Glendenning:1997wn}
\bibitem{Glendenning:1997wn} 
  N.~K.~Glendenning,
  {\it Compact stars: Nuclear physics, particle physics, and general relativity},
  Springer, New York (1997).

%\cite{Blaschke:2001uj}
\bibitem{Blaschke:2001uj} 
  D.~Blaschke, A.~Sedrakian and N.~K.~Glendenning,
  {\it Physics of neutron star interiors},
  Lecture Notes in Physics {\bf 578}, Springer, New York (2001). 
 %%CITATION = LNPHA,578,p.1;%%
  
%\cite{Haensel:2007yy}
\bibitem{Haensel:2007yy} 
  P.~Haensel, A.~Y.~Potekhin and D.~G.~Yakovlev,
  {\it Neutron stars 1: Equation of state and structure},
  Astrophysics and space science library {\bf 326}, Springer, New York (2007).
  %44 citations counted in INSPIRE as of 28 Aug 2016
    
%\cite{Blaschke:2006xt}
%\bibitem{Blaschke:2006xt} 
%  D.~Blaschke and D.~Sedrakian,
%  {\it Superdense QCD matter and compact stars},
%  Proceedings, NATO Advanced Research Workshop, Erevan, Armenia, September 27-October 4, 2003,
%  (Springer, Dordrecht, 2006)
%  (NATO science series II: Mathematics, physics and chemistry. 197)  


%\cite{Demorest:2010bx}
\bibitem{Demorest:2010bx}
  P.~Demorest, T.~Pennucci, S.~Ransom, M.~Roberts, J.~Hessels,
  %``Shapiro delay measurement of a two solar mass neutron star,''
  Nature {\bf 467}, 1081 (2010).
%  [arXiv:1010.5788 [astro-ph.HE]].

%\cite{Fonseca:2016tux}
\bibitem{Fonseca:2016tux} 
  E.~Fonseca {\it et al.},
  %``The NANOGrav Nine-year Data Set: Mass and Geometric Measurements of Binary Millisecond Pulsars,''
  arXiv:1603.00545 [astro-ph.HE].
  %%CITATION = ARXIV:1603.00545;%%
  %6 citations counted in INSPIRE as of 04 Jun 2016

%\cite{Antoniadis:2013pzd}
\bibitem{Antoniadis:2013pzd} 
  J.~Antoniadis {\it et al.},
  %``A Massive Pulsar in a Compact Relativistic Binary,''
  Science {\bf 340}, 6131 (2013).
%  [arXiv:1304.6875 [astro-ph.HE]].
  %%CITATION = ARXIV:1304.6875;%%

%\cite{Alford:2006vz}
\bibitem{Alford:2006vz} 
  M.~Alford, D.~Blaschke, A.~Drago, T.~Klahn, G.~Pagliara and J.~Schaffner-Bielich,
  %``Quark matter in compact stars?,''
  Nature {\bf 445}, E7 (2007).
%  [astro-ph/0606524].
  %%CITATION = ASTRO-PH/0606524;%%

%\cite{Klahn:2006iw}
\bibitem{Klahn:2006iw} 
  T.~Kl\"ahn, D.~Blaschke, F.~Sandin, C.~Fuchs, A.~Faessler, H.~Grigorian, G.~R\"opke and J.~Tr\"umper,
  %``Modern compact star observations and the quark matter equation of state,''
  Phys.\ Lett.\ B {\bf 654}, 170 (2007).
 % [nucl-th/0609067].
  %%CITATION = NUCL-TH/0609067;%%

%\cite{Klahn:2013kga}
\bibitem{Klahn:2013kga}
  T.~Kl\"ahn, R.~{\L}astowiecki and D.~B.~Blaschke,
  %``Implications of the measurement of pulsars with two solar masses for quark matter in compact stars and heavy-ion collisions: A NambuÐJona-Lasinio model case study,''
  Phys.\ Rev.\ D {\bf 88}, no. 8, 085001 (2013).
%  [arXiv:1307.6996].
  %%CITATION = ARXIV:1307.6996;%%
  
%\cite{Kojo:2014rca}
\bibitem{Kojo:2014rca} 
  T.~Kojo, P.~D.~Powell, Y.~Song and G.~Baym,
  %``Phenomenological QCD equation of state for massive neutron stars,''
  Phys.\ Rev.\ D {\bf 91}, no. 4, 045003 (2015).
%  [arXiv:1412.1108 [hep-ph]].
  %%CITATION = ARXIV:1412.1108;%%

%\cite{Klahn:2006ir}
\bibitem{Klahn:2006ir} 
  T.~Kl\"ahn {\it et al.},
  %``Constraints on the high-density nuclear equation of state from the phenomenology of compact stars and heavy-ion collisions,''
  Phys.\ Rev.\ C {\bf 74}, 035802 (2006).
%  doi:10.1103/PhysRevC.74.035802
%  [nucl-th/0602038].
  %%CITATION = doi:10.1103/PhysRevC.74.035802;%%

%\cite{Klahn:2011au}
\bibitem{Klahn:2011au} 
  T.~Kl\"ahn, D.~Blaschke and F.~Weber,
  %``Exploring hybrid star matter at NICA and FAIR,''
  Phys.\ Part.\ Nucl.\ Lett.\  {\bf 9}, 484 (2012).
%  doi:10.1134/S1547477112060118
%  [arXiv:1101.6061 [nucl-th]].
  %%CITATION = doi:10.1134/S1547477112060118;%%

%\cite{Hatsuda:2006ps}
\bibitem{Hatsuda:2006ps} 
  T.~Hatsuda, M.~Tachibana, N.~Yamamoto and G.~Baym,
  %``New critical point induced by the axial anomaly in dense QCD,''
  Phys.\ Rev.\ Lett.\  {\bf 97}, 122001 (2006).
%  [hep-ph/0605018].
  %%CITATION = HEP-PH/0605018;%%

%\cite{Powell:2011ig}
\bibitem{Powell:2011ig} 
  P.~D.~Powell and G.~Baym,
  %``Axial anomaly and the three-flavor Nambu--Jona-Lasinio model with confinement: Constructing the QCD phase diagram,''
  Phys.\ Rev.\ D {\bf 85}, 074003 (2012).
%  [arXiv:1111.5911 [hep-ph]].
  %%CITATION = ARXIV:1111.5911;%%
  
  
%\cite{Blaschke:2013ana}
\bibitem{Blaschke:2013ana} 
  D.~Blaschke, D.~E.~Alvarez-Castillo and S.~Benic,
  %``Mass-radius constraints for compact stars and a critical endpoint,''
  PoS CPOD {\bf 2013}, 063 (2013).
%  [arXiv:1310.3803 [nucl-th]].
  %%CITATION = ARXIV:1310.3803;%%
  
%\cite{Benic:2014jia}
\bibitem{Benic:2014jia} 
  S.~Benic, D.~Blaschke, D.~E.~Alvarez-Castillo, T.~Fischer and S.~Typel,
  %``A new quark-hadron hybrid equation of state for astrophysics - I. High-mass twin compact stars,''
  Astron.\ Astrophys.\  {\bf 577}, A40 (2015).
%  [arXiv:1411.2856 [astro-ph.HE]].
  %%CITATION = ARXIV:1411.2856;%%

\bibitem{Ayriyan:2014nua} 
  A.~Ayriyan, D.~E.~Alvarez-Castillo, D.~Blaschke, H.~Grigorian and M.~Sokolowski,
  %``New Bayesian analysis of hybrid EoS constraints with mass-radius data for compact stars,''
  Phys.\ Part.\ Nucl.\  {\bf 46}, no. 5, 854 (2015).
%  doi:10.1134/S1063779615050044
%  [arXiv:1412.8226 [astro-ph.HE]].
  %%CITATION = doi:10.1134/S1063779615050044;%%
  %7 citations counted in INSPIRE as of 04 Jun 2016
  
%\cite{Alvarez-Castillo:2016oln}
\bibitem{Alvarez-Castillo:2016oln} 
  D.~Alvarez-Castillo, A.~Ayriyan, S.~Benic, D.~Blaschke, H.~Grigorian and S.~Typel,
  %``New class of hybrid EoS and Bayesian M-R data analysis,''
  Eur.\ Phys.\ J.\ A {\bf 52}, no. 3, 69 (2016).
%  doi:10.1140/epja/i2016-16069-2
%  [arXiv:1603.03457 [nucl-th]].
  %%CITATION = doi:10.1140/epja/i2016-16069-2;%%
  
%\cite{Ohnishi:2011jv}
\bibitem{Ohnishi:2011jv} 
  A.~Ohnishi, H.~Ueda, T.~Z.~Nakano, M.~Ruggieri and K.~Sumiyoshi,
  %``Possibility of QCD critical point sweep during black hole formation,''
  Phys.\ Lett.\ B {\bf 704}, 284 (2011).
 % [arXiv:1102.3753 [nucl-th]].
  %%CITATION = ARXIV:1102.3753;%%  
  
%\cite{Bazavov:2014pvz}
\bibitem{Bazavov:2014pvz} 
  A.~Bazavov {\it et al.} [HotQCD Collaboration],
  %``Equation of state in ( 2+1 )-flavor QCD,''
  Phys.\ Rev.\ D {\bf 90}, no. 9, 094503 (2014).
%  [arXiv:1407.6387 [hep-lat]].
  %%CITATION = ARXIV:1407.6387;%%

%\cite{Borsanyi:2013bia}
\bibitem{Borsanyi:2013bia} 
  S.~Borsanyi, Z.~Fodor, C.~Hoelbling, S.~D.~Katz, S.~Krieg and K.~K.~Szabo,
  %``Full result for the QCD equation of state with 2+1 flavors,''
  Phys.\ Lett.\ B {\bf 730}, 99 (2014).
%  [arXiv:1309.5258 [hep-lat]].
  %%CITATION = ARXIV:1309.5258;%%

%\cite{Karsch:1998jm}
\bibitem{Karsch:1998jm} 
  F.~Karsch and M.~P.~Lombardo (Eds.),
  {\it QCD at finite baryon density}, 
%  Proceedings, International Workshop, Bielefeld, Germany, April 27-30, 1998,
  Nucl.\ Phys.\ A {\bf 642}, pp.1 (1998).
  %%CITATION = NUPHA,A642,pp.1;%%

%\cite{Friman:2011zz}
\bibitem{Friman:2011zz} 
  B.~Friman, C.~H\"ohne, J.~Knoll, S.~Leupold, J.~Randrup, R.~Rapp and P.~Senger,
  {\it The CBM physics book: Compressed baryonic matter in laboratory experiments},
  Lect.\ Notes Phys.\  {\bf 814}, pp. 980 (2011).
  %%CITATION = LNPHA,814,pp. 980;%%


%\cite{Nambu:1961tp}
\bibitem{Nambu:1961tp}
  Y.~Nambu, G.~Jona-Lasinio,
  %``Dynamical Model of Elementary Particles Based on an Analogy with 
  %Superconductivity. 1.,''
  Phys.\ Rev.\  {\bf 122}, 345 (1961).

%\cite{Klevansky:1992qe}
\bibitem{Klevansky:1992qe}
  S.~P.~Klevansky,
  %``The Nambu-Jona-Lasinio model of quantum chromodynamics,''
  Rev.\ Mod.\ Phys.\  {\bf 64}, 649 (1992).
  %%CITATION = RMPHA,64,649;%%

%\cite{Hatsuda:1994}
\bibitem{Hatsuda:1994}
  T.~Hatsuda, T.~Kunihiro,
  %``QCD phenomenology based on a chiral effective Lagrangian'',
  Phys. Rept. {\bf 247}, 221 (1994).

%\cite{Buballa:2003qv}
\bibitem{Buballa:2003qv} 
  M.~Buballa,
  %``NJL model analysis of dense quark matter,''
  Phys.\ Rept.\  {\bf 407}, 205 (2005).
%  [hep-ph/0402234].
  %%CITATION = HEP-PH/0402234;%%

%\cite{Fukushima:2010bq}
\bibitem{Fukushima:2010bq} 
  K.~Fukushima and T.~Hatsuda,
  %``The phase diagram of dense QCD,''
  Rept.\ Prog.\ Phys.\  {\bf 74}, 014001 (2011).
%  doi:10.1088/0034-4885/74/1/014001
%  [arXiv:1005.4814 [hep-ph]].
  %%CITATION = doi:10.1088/0034-4885/74/1/014001;%%

%\cite{Fukushima:2013rx}
\bibitem{Fukushima:2013rx} 
  K.~Fukushima and C.~Sasaki,
  %``The phase diagram of nuclear and quark matter at high baryon density,''
  Prog.\ Part.\ Nucl.\ Phys.\  {\bf 72}, 99 (2013).
%  doi:10.1016/j.ppnp.2013.05.003
%  [arXiv:1301.6377 [hep-ph]].
  %%CITATION = doi:10.1016/j.ppnp.2013.05.003;%%

%\cite{Ratti:2005jh}
\bibitem{Ratti:2005jh} 
  C.~Ratti, M.~A.~Thaler and W.~Weise,
  %``Phases of QCD: Lattice thermodynamics and a field theoretical model,''
  Phys.\ Rev.\ D {\bf 73}, 014019 (2006).
%  [hep-ph/0506234].
  %%CITATION = HEP-PH/0506234;%%

%\cite{Roessner:2006xn}
\bibitem{Roessner:2006xn} 
  S.~Roessner, C.~Ratti and W.~Weise,
  %``Polyakov loop, diquarks and the two-flavour phase diagram,''
  Phys.\ Rev.\ D {\bf 75}, 034007 (2007).
%  doi:10.1103/PhysRevD.75.034007
%  [hep-ph/0609281].
  %%CITATION = doi:10.1103/PhysRevD.75.034007;%%

%\cite{Blaschke:2010ka}
\bibitem{Blaschke:2010ka}
  D.~B.~Blaschke, F.~Sandin, V.~V.~Skokov and S.~Typel,
  %``Accessibility of Color Superconducting Quark Matter Phases in Heavy-ion Collisions,''
  Acta Phys.\ Polon.\ Supp.\  {\bf 3}, 741 (2010).
 % [arXiv:1004.4375 [hep-ph]].
  %%CITATION = ARXIV:1004.4375;%%
  %10 citations counted in INSPIRE as of 15 Apr 2015

%\cite{Schafer:1999jg}
\bibitem{Schafer:1999jg} 
  T.~Sch\"afer and F.~Wilczek,
  %``Superconductivity from perturbative one gluon exchange in high density quark matter,''
  Phys.\ Rev.\ D {\bf 60}, 114033 (1999).
%  doi:10.1103/PhysRevD.60.114033
%  [hep-ph/9906512].
  %%CITATION = doi:10.1103/PhysRevD.60.114033;%%

%\cite{Cooper:1956zz}
\bibitem{Cooper:1956zz} 
  L.~N.~Cooper,
  %``Bound electron pairs in a degenerate Fermi gas,''
  Phys.\ Rev.\  {\bf 104}, 1189 (1956).
%  doi:10.1103/PhysRev.104.1189
  %%CITATION = doi:10.1103/PhysRev.104.1189;%%
  %201 citations counted in INSPIRE as of 04 Jun 2016

%\cite{Alford:2001}
\bibitem{Alford:2001}
  M.~Alford, J.~A.~Bowers, and K.~Rajagopal,
  %``Crystalline color superconductivity'',
  Phys.\ Rev.\ {\bf D63}, 074016 (2001).

%\cite{Alford:2007xm}
\bibitem{Alford:2007xm} 
  M.~G.~Alford, A.~Schmitt, K.~Rajagopal and T.~Sch\"{a}fer,
  %``Color superconductivity in dense quark matter,''
  Rev.\ Mod.\ Phys.\  {\bf 80}, 1455 (2008).
%  doi:10.1103/RevModPhys.80.1455
%  [arXiv:0709.4635 [hep-ph]].
  %%CITATION = doi:10.1103/RevModPhys.80.1455;%%
  %613 citations counted in INSPIRE as of 04 Jun 2016

\bibitem{Anglani:2013gfu}
  R.~Anglani, R.~Casalbuoni, M.~Ciminale, N.~Ippolito, R.~Gatto, M.~Mannarelli and M.~Ruggieri,
  %``Crystalline color superconductors,''
  Rev.\ Mod.\ Phys.\  {\bf 86}, 509 (2014).
%  doi:10.1103/RevModPhys.86.509
%  [arXiv:1302.4264 [hep-ph]].
  %%CITATION = doi:10.1103/RevModPhys.86.509;%%
  %38 citations counted in INSPIRE as of 04 Jun 2016

%\cite{GomezDumm:2008sk}
\bibitem{GomezDumm:2008sk}
  D.~Gomez Dumm, D.~B.~Blaschke, A.~G.~Grunfeld, N.~N.~Scoccola,
  %``Color neutrality effects in the phase diagram of the PNJL model,''
  Phys.\ Rev.\  {\bf D78}, 114021 (2008).
  %[arXiv:0807.1660 [hep-ph]].

%\cite{Powell:2013cq}
\bibitem{Powell:2013cq} 
  P.~D.~Powell and G.~Baym,
  %``Asymmetric pairing of realistic mass quarks and color neutrality in the PolyakovÐNambuÐJona-Lasinio model of QCD,''
  Phys.\ Rev.\ D {\bf 88}, no. 1, 014012 (2013).
%  [arXiv:1302.0416 [hep-ph]].
  %%CITATION = ARXIV:1302.0416;%%

%\cite{Blaschke:2005uj}
\bibitem{Blaschke:2005uj}
  D.~Blaschke, S.~Fredriksson, H.~Grigorian, A.~M.~\"Oztas, F.~Sandin,
  %``The Phase diagram of three-flavor quark matter under compact star 
  %constraints,''
  Phys.\ Rev.\  {\bf D72}, 065020 (2005).
  %[hep-ph/0503194].

%\cite{Ruester:2005jc}
\bibitem{Ruester:2005jc} 
  S.~B.~Ruester, V.~Werth, M.~Buballa, I.~A.~Shovkovy and D.~H.~Rischke,
  %``The Phase diagram of neutral quark matter: Self-consistent treatment of quark masses,''
  Phys.\ Rev.\ D {\bf 72}, 034004 (2005).
%  doi:10.1103/PhysRevD.72.034004
%  [hep-ph/0503184].
  %%CITATION = doi:10.1103/PhysRevD.72.034004;%%

%\cite{Ruester:2005ib}
\bibitem{Ruester:2005ib} 
  S.~B.~Ruester, V.~Werth, M.~Buballa, I.~A.~Shovkovy and D.~H.~Rischke,
  %``The Phase diagram of neutral quark matter: The Effect of neutrino trapping,''
  Phys.\ Rev.\ D {\bf 73}, 034025 (2006).
%  doi:10.1103/PhysRevD.73.034025
%  [hep-ph/0509073].
  %%CITATION = doi:10.1103/PhysRevD.73.034025;%%

%\cite{Sandin:2007zr}
\bibitem{Sandin:2007zr} 
  F.~Sandin and D.~Blaschke,
  %``The quark core of protoneutron stars in the phase diagram of quark matter,''
  Phys.\ Rev.\ D {\bf 75}, 125013 (2007).
%  doi:10.1103/PhysRevD.75.125013
%  [astro-ph/0701772].
  %%CITATION = doi:10.1103/PhysRevD.75.125013;%%

%\cite{Blaschke:2010vj}
\bibitem{Blaschke:2010vj} 
  D.~Blaschke, J.~Berdermann and R.~Lastowiecki,
  %``Hybrid neutron stars based on a modified PNJL model,''
  Prog.\ Theor.\ Phys.\ Suppl.\  {\bf 186}, 81 (2010).
%  doi:10.1143/PTPS.186.81
%  [arXiv:1009.1181 [nucl-th]].
  %%CITATION = doi:10.1143/PTPS.186.81;%%

%\cite{Grigorian:2006qe}
\bibitem{Grigorian:2006qe} 
  H.~Grigorian,
  %``Parametrization of a nonlocal, chiral quark model in the instantaneous three-flavor case: Basic formulas and tables,''
  Phys.\ Part.\ Nucl.\ Lett.\  {\bf 4}, 223 (2007).
%  doi:10.1134/S1547477107030041
%  [hep-ph/0602238].
  %%CITATION = doi:10.1134/S1547477107030041;%%

\bibitem{Sandin}
For details of calculation and an online parametrization solving tool see webpage by F.~Sandin at 
{\tt http://3fcs.pendicular.net}

  
%\cite{Buballa:2005bv}
\bibitem{Buballa:2005bv} 
  M.~Buballa and I.~A.~Shovkovy,
  %``A Note on color neutrality in NJL-type models,''
  Phys.\ Rev.\ D {\bf 72}, 097501 (2005).
%  doi:10.1103/PhysRevD.72.097501
%  [hep-ph/0508197].
  %%CITATION = doi:10.1103/PhysRevD.72.097501;%%
  
%\cite{He:2005jq}
\bibitem{He:2005jq} 
  L.~Y.~He, M.~Jin and P.~F.~Zhuang,
  %``On the ground state of two flavor color superconductor,''
  Chin.\ Phys.\ Lett.\  {\bf 23}, 564 (2006).
%  doi:10.1088/0256-307X/23/3/011
%  [hep-ph/0505061].
  %%CITATION = doi:10.1088/0256-307X/23/3/011;%%

%\cite{Blaschke:2005km}
\bibitem{Blaschke:2005km} 
  D.~Blaschke, D.~Gomez Dumm, A.~G.~Grunfeld and N.~N.~Scoccola,
  %``Color neutral ground state of 2SC quark matter,''
  hep-ph/0507271.
  %%CITATION = HEP-PH/0507271;%%
  
%\cite{Abuki:2009dt}
\bibitem{Abuki:2009dt} 
  H.~Abuki and K.~Fukushima,
  %``Gauge dynamics in the PNJL model: Color neutrality and Casimir scaling,''
  Phys.\ Lett.\ B {\bf 676}, 57 (2009).
%  doi:10.1016/j.physletb.2009.04.078
%  [arXiv:0901.4821 [hep-ph]].
  %%CITATION = doi:10.1016/j.physletb.2009.04.078;%%
  
%\cite{Petran:2013qla}
\bibitem{Petran:2013qla} 
  M.~Petran and J.~Rafelski,
  %``Universal hadronization condition in heavy ion collisions at $\sqrt{s_\mathrm{NN}}= 62$ GeV and at $\sqrt{s_\mathrm{NN}}=2.76$ TeV,''
  Phys.\ Rev.\ C {\bf 88}, 021901(R) (2013).
%  doi:10.1103/PhysRevC.88.021901
%  [arXiv:1303.0913 [hep-ph]].
  %%CITATION = doi:10.1103/PhysRevC.88.021901;%%
  %23 citations counted in INSPIRE as of 12 Jul 2016

%\cite{Ivanov:2016hes}
\bibitem{Ivanov:2016hes} 
  Y.~B.~Ivanov and A.~A.~Soldatov,
  %``Entropy Production and Effective Viscosity in Heavy-Ion Collisions,''
  arXiv:1605.02476 [nucl-th].
  %%CITATION = ARXIV:1605.02476;%%

%\cite{Ayriyan:2016pat}
\bibitem{Ayriyan:2016pat} 
  A.~Ayriyan, D.~Blaschke and R.~Lastowiecki,
  %``Phase diagram of the three-flavor color superconducting PNJL model,''
  J.\ Phys.\ Conf.\ Ser.\  {\bf 668}, no. 1, 012101 (2016).
%  doi:10.1088/1742-6596/668/1/012101
  %%CITATION = doi:10.1088/1742-6596/668/1/012101;%%

\end{thebibliography}
\end{document}